\documentclass{aastex}
\usepackage{spr-astr-addons}
\usepackage{url}

\usepackage{floatrow}
\usepackage{multirow}

\RequirePackage{color}

\newcommand{\emaila}

\begin{document}

\title{Photon in the Earth-ionosphere cavity: Schumann resonances}

\shorttitle{Photon in the Earth-ionosphere cavity: Schumann resonances}
\shortauthors{Y. Sucu and C. Tekincay}

\author{{Y. Sucu and C. Tekincay}}
\affil{Department of Physics, Faculty of Sciences, Akdeniz University, Antalya, Turkey
\emaila{ysucu@akdeniz.edu.tr and ctekincay@akdeniz.edu.tr}}

\begin{abstract}
We study a quantum analogy of Schumann resonances by solving massless and massive spin-1 particle equations derived from the Zitterbewegung model in an annular cavity background with poorly conducting walls. We also show that the massless case and the massive case in the $m_0^2 \rightarrow 0$ limit are compatible with Maxwell's electromagnetic theory. Furthermore, from the massive case, we predict an upper limit for the mass of the photon of 1.3x$10^{-50}$kg. The bound on the mass of the photon is compatible with the current limit in the literature.
\end{abstract}


\section{Introduction}
\label{sec:1}
\hspace*{0.5cm} Electromagnetic discharges in the Earth-ionosphere cavity, which theoretically assumed concentric spheres with perfectly conducting walls, generate extremely low frequency (ELF) noises, e.g. 10.6, 18.3, 25.9, 33.5 Hz \citep{Madden1965}, known as Schumann resonances (SR) \citep{schumann}. Since this phenomenon has been noticed it has received increased attention and has found many applications in various areas of natural sciences  \citep{balser,galejs,Madden1965,sentman,Williams,Mushtak2002,cherry2002,cherry2003,Yair,kozlowski2015schumann,price,satori,gazquez,shvets,Toledo,Alabdulgader}.  Recently, it is also discussed whether the detection of gravitational waves are affected by the electromagnetic fields from SR \citep{Coughlin2016,Kowalska2017,silagadze2018}.  

In classical aspect of physics, the ELF noises are electromagnetic radiation and they can be measured by deviations in the electric field or magnetic field which theoretically satisfy the Maxwell equations \citep{sanfui,Sanfui,palangio}. On the other hand, electromagnetic radiation is quantized by $h\nu$ energized photons \citep{planck} and it can be measured by absorption of photons which again theoretically must obey Maxwell equations. Accordingly, there is a very early theoretical study done by \cite{Kroll1971} on determining SR and the quality factor from the massless photon case and an upper limit on the photon rest mass from the massive photon case by using the Klein-Gordon equation. In addition, there are plenty of studies that try to set an upper limit on the photon rest-mass by experimental, observational and theoretical methods, see also \cite{tu,goldhaber}. \cite{Kroll1971} use the \textit{relativistic particle in a box} formula by adding a general complex parameter including different heights of the ionosphere layers which is derived from applying boundary conditions on the solutions of Maxwell equations. There is a contradiction here because they think that the Klein Gordon particle (spin-0) and photon (spin-1) are the same relativistic quantum mechanical particles, and at the same time they solve Maxwell equations which do not include quantum mechanical corrections such as the spin quantum number. However, we have massless and massive spin-1 relativistic quantum mechanical equations to describe photons which have no contradiction in this physical sense.  Furthermore, the equivalency of these equations to the spinor form of the Maxwell equations, in which the spinor form of the Maxwell fields are defined as $\vec{E}+i\vec{B}$, is discussed by many authors \citep{jena, sucu2002, sucu2005}. Therefore, the quantum analogy of SR can be investigated by using the massless and the massive spin-1 particle equations and we may set an upper limit on the photon mass so that gauge invariance will be safe \citep{jacksongauge}. 

The mathematical physics used in this study is based on the excited states of the Zitterbewegung (trembling motion) model \citep{barut}. This model presents a pure spin-1 particle equation (either massive or massless) and corresponds to the spin-1 part of the Duffin-Kemmer-Petiau (DKP) equation \citep{duffin,kemmer,petiau} in the flat spacetime. Then, it is generalized to the curved spacetime to investigate the massive spin-1 particle (vector boson) creation in the expanding universe and it was shown that the results are compatiple with Maxwell's electromagnetic theory in the $m_0^2 \rightarrow 0$ limit \citep{sucu2002, sucu2005}. Also, the symmetry and the integrability properties of this model are carried out \citep{sucu2012}. 

The massless case treated by the equation for the spin-1 particle is a toy model of the Zitterbewegung and it was studied by \cite{unal}. Then, it is studied on the massless spin-1 particle creation in the Robertson-Walker spacetime background \citep{sucu2002}. In addition, the DKP equation in the $m_0 \rightarrow 0$ limit which describes the photon \citep{jena} and massless DKP fields \citep{casana} are also studied. All these studies are based on the equivalence of the massless spin-1 particle equation and the Maxwell equations. After these, some recent studies on the spin-1 particle are quantum dynamics of vector bosons \citep{castro, oliveira}, tunneling properties of vector bosons \citep{gecim2017a}, the generalized uncertainty principle (GUP) effect on vector bosons \citep{gecim2017b}, generalized bosonic oscillator via minimal length uncertainty \citep{falek}, resonance frequencies of the photon in a cylindrical resonant cavity \citep{tekincay} and the pair production with the Noether charge in $2+1$ dimensional spacetime backgrounds \citep{dernek2017}. All these studies are important in consequence of the description of the quantum electrodynamical behaviour of the photon. Therefore, in the study, we consider this a useful application for the photon description.

The paper is organized as follows: in the following Sections, the massless and the massive spin-1 particle equations are respectively solved in an annular resonant cavity background with poorly conducting walls. We show that the resonance energy expressions obtained from the solutions are compatible with both each other and the Maxwell's electromagnetic theory in the $m_0^2 \rightarrow 0$ limit. Finally, we compare the first six modes of SR with the experimental results and constrain the photon mass.

\section{The Massless spin-1 particle in an annular resonant cavity}
\label{sec:2}
\hspace*{0.5cm} 
The Maxwell equations perfectly describe the nature of light in terms of the electromagnetic fields in the context of classical physics. To be understood the quantum nature of the light, either the electromagnetic fields are quantized in the context of quantum field theory or the equivalence of the spin-1 equations to the Maxwell field equations are derived in the classical limit \citep{Oppenheimer, Nelson, Good, Weinberg, Mignani, Gianatto, Dvoeglazov, Gersten, Kobe, Leonhardt2000}. In this context,  the covariant form of the massless spin-1 particle equation in a curved spacetime is given as \citep{sucu2002}:
\begin{equation}\label{masslesseq}
i\hbar\Sigma^{\mu}(x)[\partial_{\mu}+\Gamma_\mu(x)\otimes I + I \otimes \Gamma_\mu(x) ] \psi(x)=0 ,
\end{equation}
where $\Sigma^{\mu}(x)=\sigma^\mu(x)\otimes I + I \otimes \sigma^\mu(x)$, $\psi(x)$ is $4\times 1$ spinor and $\hbar$ is reduced Planck constant. Here, the model construction of the spin-1 particle is done by extending the space of spin-1/2. So, we need the spin connection for spin-1/2, $\Gamma_\mu(x)$, is defined as follows \citep{sucu2002}:
\begin{equation}
\Gamma_\mu(x)=-\frac{1}{8}\big[ \sigma^\nu(x),\sigma_{\mu;\nu}(x)\big] ,
\end{equation}
where $\sigma^\mu(x)$ represents the Pauli matrices in the general coordinate frame which transform as 
\begin{equation}
\sigma^\mu(x)=e^{\mu}_{\tilde{a}}(x)\sigma^{\tilde{a}} ,
\end{equation}
where $e^{\mu}_{\tilde{a}}(x)$ is the tetrad which satisfies the following relation
\begin{equation}\label{tetradrelation}
g^{\mu\nu}=e^{\mu}_{\tilde{a}}(x)e^{\nu}_{\tilde{b}}(x)\eta^{\tilde{a}\tilde{b}} .
\end{equation}
The metric tensor and the tetrad of the annular resonant cavity with a constant factor, $n$, can be written as:
\begin{equation}
g_{\mu\nu}=diag[-1,-r^2,-r^2 sin^2\theta,c^2 n^2], \label{metric}
\end{equation}
\begin{equation}
e^{\mu}_{\tilde{a}}(x)=diag\Big [1,\frac{1}{r},\frac{1}{r sin\theta},\frac{1}{c n}\Big]. \label{tetrad}
\end{equation}
where $c$ is the speed of light in vacuum, $n=1-\frac{1}{Q}$ and $Q$ is a quality factor of the cavity \citep{galejs}. Then, we can take the rotated Pauli matrices as follows \citep{sucu2002}:
\begin{equation}\label{Pauli}
\sigma^1(x) = -\sigma_3,\:\: \sigma^2(x) = -\frac{1}{r} \sigma_1, \:\: \sigma^3(x) = - \frac{1}{rsin\theta}\sigma_2 ,
\end{equation}
and $\sigma^4(x)=\sigma_0/n$ where $\sigma_0$, $2\times 2$ unit matrix.

Therefore, the non-zero spin connections in this context are indicated in Eq. (\ref{nonzeromassless}).
\begin{equation}\label{nonzeromassless}
\Gamma_2(x)=-\frac{i}{2}\sigma_2, \:\:\Gamma_3(x)=\frac{isin\theta}{2}\sigma_1-\frac{icos\theta}{2}\sigma_3 .
\end{equation}
Using the non-zero spin connections, the massless spin-1 particle equation in the annular resonant cavity can then be written as:
\begin{equation}\label{masslesseqstep}
\big\{ \Sigma_3  (\partial_r+\frac{1}{r})-\frac{1}{r}[ \Sigma_{+}\partial_{+}-\Sigma_{-}\partial_{-}]
-\frac{2}{n} I\otimes I\partial_t\big\}\psi=0 ,
\end{equation}
where the angular part of the Eq. (\ref{masslesseqstep}) is defined in terms of the ladder operator, $\partial_{\mp}$ \citep{sucu2002}, 
\begin{equation}\label{angularpart1}
\Sigma_{+}\partial_{+}-\Sigma_{-}\partial_{-}=-\Sigma_1 \partial_\theta +i\Sigma_2 \big(\frac{i\partial_\phi}{sin\theta}+\frac{cot\theta}{2}\Sigma_3\big) .
\end{equation}
By the separation of variables method, the $4\times 1$ spinor for the annular resonant cavity is defined as:
\begin{equation}\label{massless spinor}
\psi= \frac{e^{-iE t/\hbar}}{r}
\left( \begin{array}{c}
R_{+}\\R_{0}\\R_{0}\\R_{-} 
\end{array} \right)  D^{j}_{\lambda,m},
\end{equation}
where $E$ is the energy of the spin-1 particle, $R(r)$ is the radial function, $D^{j}_{\lambda,m}(\theta,\phi)$ is the eigenfunction of the $\partial \pm$ ladder operator and SR depends on the $j$ eigenvalue of the $\partial \pm$ operator.  Also, $D^{j}_{\lambda,m}(\theta,\phi)$ is called the Wigner matrix and defines the irreducible representation of pure rotation group SU(2) \citep{wigner}. 

The eigenvalues of Eq. (\ref{angularpart1}) are given by \citep{sucu2002}: 
\begin{equation}\label{angularpart2}
\int sin\theta d\theta d\phi D^{*j}_{\lambda,m}[\Sigma_{+}\partial_{+}-\Sigma_{-}\partial_{-}]D^{j}_{\lambda,m}=i\sqrt{j(j+1)}\Sigma_2  ,
\end{equation}
where the quantum number of total angular momentum, $j$, contains the quantum number of orbital angular momentum, $l$, and the quantum number of spin angular momentum, $s$, as follows:
\begin{equation}\label{totalj}
j=\left \{ \begin{array}{l}l+s \\ l \\ l-s \end{array} \right.
\end{equation}
where $l=0,1,2,...$ and $s=1$ are different from the spin-0 case in the study of \cite{Kroll1971}.

Using the eigenvalues of the angular momentum eigenfunction in Eq. (\ref{angularpart2}), the massless spin-1 equation in Eq. (\ref{masslesseqstep}) gives the following three expressions:
\begin{equation}
\Big[\frac{d}{dr}+\frac{iE}{\hbar cn} \Big]R_{+}=\frac{\sqrt{j(j+1)}}{r}R_0 ,
\end{equation}
\begin{equation}
\Big[\frac{d}{dr}-\frac{iE}{\hbar cn} \Big]R_{-}=\frac{\sqrt{j(j+1)}}{r}R_0 ,
\end{equation}
\begin{equation}\label{constants}
\frac{\sqrt{j(j+1)}}{r}[R_{+}-R_{-}]=-\frac{iE}{2\hbar cn}R_{0} .
\end{equation}
Adding and subtracting these equations, we find the spherical Bessel differential equation for $(R_+ - R_-)$ which also can be derived from the Maxwell equations but would have no contribution of the spin quantum number \citep{abramowitz1964}:
\begin{equation}\label{masslesseq0}
\Bigg[  \frac{d^2}{dr^2}+\frac{E^2}{\hbar^2 c^2 n^2}-\frac{j(j+1)}{r^2} \Bigg] (R_{+}-R_{-})=0.
\end{equation}
The Bessel differential equation is equivalent to the spinor form of the Maxwell equations and the only difference comes from the total angular momentum. Then, it yields the following solution:
\begin{equation}\label{masslesssol1}
R_{+}-R_{-}=N (r\textnormal{j}_j),
\end{equation}
where $\textnormal{j}_j(\frac{E r}{\hbar cn})$ is the spherical Bessel function for integer $j$. We eliminate the other independent solution, the spherical Neumann function, because it goes to $-\infty$ while its argument goes to zero and we do not want infinite spinor components so that the probability current remains finite. 

The other solutions can be found by using Eq. (\ref{masslesssol1}) as follows:
\begin{equation}\label{masslesssol2}
R_{0}=N\frac{i\hbar cn }{E }\sqrt{j(j+1)}  \textnormal{j}_j ,
\end{equation}
\begin{equation}\label{masslesssol3}
R_{+}+R_{-}= N\frac{i\hbar cn }{E }\frac{d(r\textnormal{j}_j)}{dr}.
\end{equation}
\begin{table*}[h]\label{t1}
\centering
\caption{Calculated and measured results for the SR. The data shown below are taken from the studies of \cite{Boldi2018}$^{\ast}$, \cite{nicko}$^{\dagger}$, \cite{Kroll1971}$^{\ddagger}$ respectively.}
\begin{tabular}{cccccc|cc|cc|cc}
\hline 
\multicolumn{6}{c|}{\bf Theoretical }
&\multicolumn{2}{c|}{\bf Experimental$^{\ast}$}
&\multicolumn{2}{c|}{\bf Experimental$^{\dagger}$}
&\multicolumn{2}{c}{\bf Theoretical$^{\ddagger}$}\\
\multicolumn{1}{c}{$l$ }
&\multicolumn{1}{c}{$j$ } 
&\multicolumn{1}{c}{$f_j$ }
&\multicolumn{1}{c}{$\epsilon_j$ }
&\multicolumn{1}{c}{$\overline{Q_j}$}  
&\multicolumn{1}{c|}{$\nu_j$} 
&\multicolumn{1}{c}{$Q$}
&\multicolumn{1}{c|}{$\nu$}  
&\multicolumn{1}{c}{$Q$}
&\multicolumn{1}{c|}{$\nu$}
&\multicolumn{1}{c}{$Q$}
&\multicolumn{1}{c}{$\nu$}\\ \hline 
0& &    &    &     &    &	  &    &     &     &     &     \\
1&1&10.6&0.28& 3.66& 7.7& 2.79& 7.8& 4.0 &  7.8&  4.8& 7.0 \\
2& &    &    &     &    &     &    &     &     &     &     \\  \hline
1& &    &    &     &    &	  &    &     &     &     &     \\
2&2&18.3&0.16& 4.76&14.5& 5.35&14.2& 4.5 & 13.9&  5.9&13.0 \\
3& &    &    &     &    &     &    &     &     &     &     \\  \hline
2& &    &    &     &    &     &    &     &     &     &     \\
3&3&25.9&0.15& 5.16&20.9& 5.82&20.1& 5.0 & 20.0&  5.8&19.1 \\
4& &    &    &     &    &     &    &     &     &     &     \\  \hline
3& &    &    &     &    &     &    &     &     &     &     \\
4&4&33.5&0.19& 5.18&27.0& 5.81&26.5& 5.5 & 26.0&  6.9&25.7 \\
5& &    &    &     &    &     &    &     &     &     &     \\  \hline 
4& &    &    &     &    &     &    &     &     &     &     \\
5&5&41.0&0.30& 4.93&32.7& 5.22&32.2& 6.0 & 32.0&  -  &  -  \\
6& &    &    &     &    &  	  &    &     &     &     &     \\  \hline 
5& &    &    &     &    &     &    &     &     &     &     \\
6&6&48.5&0.30& 5.09&39.0& 5.82&38.6&  -  &  -  &  -  &  -  \\
7& &    &    &     &    &     &    &     &     &     &     \\  \hline 
\end{tabular}
\end{table*} 

The boundary condition defining the SR cavity is given by \cite{jackson}
\begin{equation}\label{boundary}
 \frac{d(r\textnormal{j}_j)}{dr}=0,\quad \textnormal{for} \:\: r=a \:\: \textnormal{and} \:\: r=b ,
\end{equation}
where $a$ is radius of the Earth ($\approx 6370$ km) and $b$ is the approximate radius of the ionosphere ($\approx 6445$ km) \citep{bliokh}. We know that the height of the ionosphere changes diurnally and seasonally due to the changes in the electrical conductivity of the ionosphere \citep{galejs2}. However, this problem is already studied by \cite{Greifinger2007}. Here, we do not get into the profile of the ionosphere, instead we calculate SR by taking an approximate radius of it since the height up to $\approx$100 km can be safely neglected when we compare it with the radius of Earth as the assumption $\frac{b-a}{a} \ll 1$. With this assumption, applying Eq. (\ref{boundary}) to Eq. (\ref{masslesseq0}), we find the approximate SR energies ($E=h\nu$) as follows:
\begin{equation}\label{masslessreso}
E_{j} \approx \frac{hc }{2\pi b} \sqrt{j(j+1)} \Big(1-\frac{1}{Q}\Big),
\end{equation}
Here the total angular momentum numbers take values for the spin-1 particle as $j=1$ when only $l=0,1,2$; $j=2$ when only $l=1,2,3$; $j=3$ when only $l=2,3,4$ and so on. Therefore, the photon spin polarizations allow us to choose the various orbital angular momentum numbers such that the total angular momentum quantum number, $j$, has same value. 

To determine the average quality factor value ($\overline{Q_j}$), we extract $Q_j$ from Eq. (\ref{masslessreso}) for each resonance frequency as follows: 
\begin{equation}
Q_j=\frac{f_j}{f_j-\nu_j}
\end{equation}
where $\nu_j$ is the observed resonance frequency and $f_j=c \sqrt{j(j+1)}/2\pi b $ is the maximum possible resonance frequency called \textit{natural resonance frequency of the cavity}. Then we use the mean value theorem for $Q_j$ by regularizing the integral for each total angular momentum quantum number ($j$):
\begin{equation}
\overline{Q_j}= \int_{0}^{f_j} \frac{d\nu_j}{f_j-\nu_j+\epsilon_j}
\end{equation}
where $\epsilon_j$ is a regularization parameter. Then the integral yields
\begin{equation}\label{quality}
\overline{Q_j}=ln(1+\frac{f_j}{\epsilon_j})
\end{equation} 
and we recover the perfectly conducting walls concentric cavity case ($Q\rightarrow \infty$) in the $\epsilon_j \rightarrow 0$ limit.
Now, our resonance frequency relation in Eq. (\ref{masslessreso}) becomes
\begin{equation}\label{lastnu}
v_j\approx f_j\left( 1- \frac{1}{ln(1+f_j/\epsilon_j)}\right)
\end{equation}
In Table 1, the frequency-dependent averaged quality factor $\overline{Q}_j$ and the resonance frequency $\nu_j$ are calculated from Eq. (\ref{quality}) and  Eq. (\ref{lastnu}) by using the regularization parameter ($\epsilon_j$). 

\section{The Massive spin-1 particle in a spherical resonant cavity}
\label{sec:3}
\hspace*{0.5cm} 
We now concentrate on finding an upper limit on the photon (spin-1 particle) rest mass. To do so, we use the covariant form of the massive spin-1 particle equation in a curved spacetime which is given by \citep{sucu2005}:
\begin{equation}\label{massiveequ}
\big\{i\beta^{\mu}(x)[\partial_{\mu}-\Gamma_\mu(x)\otimes I - I \otimes \Gamma_\mu(x) ] - M\big\}_{\alpha\beta,\gamma			 	 \delta}  \Psi_{\gamma\delta}(x)=0,
\end{equation}
where  $\beta^{\mu}(x)=[\gamma_\mu(x)\otimes I + I \otimes \gamma_\mu(x)]/2$ represents the Kemmer matrices and $\Psi_{\gamma\delta}(x)$ is the $16\times 1$ symmetric spinor \citep{sucu2005}. Also $M=m_0 c/\hbar$ is the inverse Compton wavelength of the spin-1 particle where its mass is $m_0$ and $c$ is the speed of light in vacuum. The spin connection for spin-1/2, $\Gamma_\mu(x)$, is defined as \citep{sucu2002}:
\begin{equation}
\Gamma_\mu(x)=-\frac{1}{8}\big[ \gamma^\nu(x),\gamma_{\mu;\nu}(x)\big],
\end{equation}
where $\gamma^\mu(x)$ represents the Dirac matrices in the general coordinate frame which are transformed by means of a tetrad, $e^{\mu}_{\tilde{a}}$, as follows
\begin{equation}
\gamma^\mu(x)=e^{\mu}_{\tilde{a}}(x)\gamma^{\tilde{a}}.
\end{equation}
They also satisfy the anti-commutation relation
\begin{equation}\label{anticom}
\{\gamma^\mu(x),\gamma^\nu(x)\}=2g^{\mu\nu}.
\end{equation}
The metric tensor and the tetrad of the spherical resonant cavity background are defined in Eqs. (\ref{metric}-\ref{tetrad}). Therefore, the non-zero spin connections yield the following expressions:
\begin{align}\label{nonzeromassive}
&\Gamma_2(x)=\frac{i}{2} 
\left(\begin{array}{c c}
\sigma_2 & 0 \\ 0 & \sigma_2
\end{array}\right), \nonumber \\
&\Gamma_3(x)=\frac{icos\theta}{2}
\left(\begin{array}{c c}
\sigma_3 & 0 \\ 0 & \sigma_3
\end{array}\right)
-\frac{isin\theta}{2}
\left(\begin{array}{c c}
\sigma_1 & 0 \\ 0 & \sigma_1
\end{array}\right).
\end{align}
Using the non-zero spin connections in Eq. (\ref{nonzeromassive}), the equation for massive spin-1 particles for the spherical resonant cavity can then be written as follows:
\begin{equation}\label{massiveequ1}
\left( \begin{array}{c c c c}
\frac{2}{n}\partial_t+2iM & A & B & 0 \\
-A & 2iM & 0 & B\\
-B & 0 & 2iM & A\\
0 & -B & -A & -\frac{2}{n}\partial_t+2iM
\end{array} \right)\Psi=0,
\end{equation}
where $A$ and $B$ are
\begin{equation}\label{A}
A=I\otimes \vec{\sigma}\cdot \vec{\nabla}+\frac{i}{2r}[I\otimes\sigma_1 \Sigma_2 + I\otimes \sigma_2(cot\theta \Sigma_3-\Sigma_1)],
\end{equation}
\begin{equation}\label{B}
B= \vec{\sigma} \otimes I \cdot \vec{\nabla}+\frac{i}{2r}[\sigma_1 \otimes I \Sigma_2 + \sigma_2 \otimes I (cot\theta \Sigma_3-\Sigma_1)].
\end{equation}
By the separation of variables method, the $16\times 1$ symmetric spinor for the spherical resonant cavity is defined as follows:
\begin{equation}\label{massivespinor1}
\Psi=
\left( \begin{array}{c}
\psi_1 \\ \psi_2  \\ \psi_3 \\ \psi_4
\end{array} \right)
= \frac{e^{-iE t/\hbar}}{ r} 
\left( \begin{array}{c}
R_{1}\\ R_{2}\\ R_{3}\\ R_{4}
\end{array} \right) D^{j}_{\lambda,m},
\end{equation}
where $D^{j}_{\lambda,m}$ is Wigner matrix \citep{wigner} and $R$ is the radial function and its explicit form is
\begin{equation}\label{massivespinorradial1}
\begin{array}{c c}
R_1=\left( \begin{array}{c}
R_{1+} \\ R_{10}  \\ R_{10} \\ R_{1-}
\end{array} \right),& \hspace*{0.5cm}
R_2=\left( \begin{array}{c}
R_{2+} \\ R_{20}  \\ R_{2\tilde{0}} \\ R_{2-}
\end{array} \right)\\
\end{array}  ,
\end{equation}
\begin{equation}\label{massivespinorradial2}
\begin{array}{c c}
R_3=\left( \begin{array}{c}
R_{2+} \\ R_{2\tilde{0}}  \\ R_{20} \\ R_{2-}
\end{array} \right),&\hspace*{0.5cm}
R_4=\left( \begin{array}{c}
R_{4+} \\ R_{40}  \\ R_{40} \\ R_{4-}
\end{array} \right)
\end{array}.
\end{equation}
Using the $16 \times 1$ symmetric spinor, Eq. (\ref{massiveequ1}) becomes
\begin{equation}\label{M1}
-\frac{2iE}{\hbar cn}(\psi_1-\psi_4)+2iM(\psi_1+\psi_4)+[A-B](\psi_2-\psi_3)=0 ,
\end{equation}
\begin{equation} \label{M2}
	-\frac{2iE}{\hbar cn}(\psi_1+\psi_4)+2iM(\psi_1-\psi_4)+[A+B](\psi_2-\psi_3)=0,
\end{equation}
\begin{equation}\label{M3}
2iM(\psi_2+\psi_3)-[A+B](\psi_1-\psi_4)=0,
\end{equation}
\begin{equation}\label{M4}
2iM(\psi_2-\psi_3)-[A-B](\psi_1+\psi_4)=0,
\end{equation}
where $A+B$ and $A-B$ are defined respectively:
\begin{equation}\label{A+B} 
A+B=-\Sigma_3 (\partial_r+\frac{1}{r}) + \frac{1}{r}[ \Sigma_{+}\partial_{+}-\Sigma_{-}\partial_{-}],
\end{equation}
\begin{equation}\label{A-B}
A-B= \overline{\Sigma}_3 (\partial_r+\frac{(\Sigma_2)^2}{2r}) - \frac{1}{r}[ \overline{\Sigma}_{+}\partial_{+}-\overline{\Sigma}_{-}\partial_{-}],
\end{equation}
where, as an abbreviation, we define $\overline{\Sigma}_b = \sigma_b \otimes I -I \otimes \sigma_b$ and $b=1,2,3$.
Therefore, the eigenvalues of the angular momentum eigenfunction in Eq. (\ref{A-B}) are found as follows:
\begin{equation}\label{angularpart3}
\int sin\theta d\theta d\phi D^{*j}_{\lambda,m}[\overline{\Sigma}_{+}\partial_{+}-\overline{\Sigma}_{-}\partial_{-}]D^{j}_{\lambda,m}=i\sqrt{j(j+1)}\overline{\Sigma}_2 ,
\end{equation}
just as in Eq. (\ref{angularpart1}).

Adding and subtracting of Eqs. (\ref{M1}-\ref{M4}), we find the transverse ($\pm$ helicity) states
\begin{align}
&(R_1+R_4)_{\pm}=\frac{i}{u^2} \Big\{ -M\frac{\sqrt{j(j+1)}}{r}(R_{20}-R_{2\tilde{0}}) \nonumber \\
&- \frac{E}{\hbar cn} \Big[\mp 2\frac{d R_{2\pm}}{dr}\pm \frac{\sqrt{j(j+1)}}{r}(R_{20}+R_{2\tilde{0}}) \Big] \Big\},
\end{align}
\begin{align}
&(R_1-R_4)_{\pm}=\frac{i}{u^2} \Big\{-\frac{E}{\hbar cn}\frac{\sqrt{j(j+1)}}{r}(R_{20}-R_{2\tilde{0}}) \nonumber \\ &+M \Big[\pm 2\frac{d R_{2\pm}}{dr}\mp \frac{\sqrt{j(j+1)}}{r}(R_{20}+R_{2\tilde{0}}) \Big]\Big\},
\end{align}

\begin{equation}\label{eq1}
R_{2+}+R_{2-}=\frac{1}{\sqrt{j(j+1)}}\frac{d[r(R_{20}+R_{2\tilde{0}})]}{dr},
\end{equation}
\begin{equation}\label{eq2}
\Big\{\frac{d^2}{dr^2}+u^2 -\frac{j(j+1)}{r^2} \Big\}(R_{2+}-R_{2-})=0,
\end{equation}
and the longitudinal (zero helicity) states
\begin{eqnarray}
(R_1+R_4)_{0}&=&\frac{i}{u^2} \Big\{ -M (\frac{d}{dr}-\frac{1}{r})(R_{20}-R_{2\tilde{0}})\nonumber \\
&+&\frac{E}{\hbar c}\frac{\sqrt{j(j+1)}}{r}(R_{2+}-R_{2-})\Big\},
\end{eqnarray}
\begin{eqnarray}
(R_1-R_4)_{0}&=&\frac{i}{u^2} \Big\{ M\frac{\sqrt{j(j+1)}}{r}(R_{2+}-R_{2-}) \nonumber \\
&-&\frac{E}{\hbar c} (\frac{d}{dr}-\frac{1}{r})(R_{20}-R_{2\tilde{0}}) \Big\},
\end{eqnarray}
\begin{equation}\label{eq3}
\Big\{\frac{d^2}{dr^2}+u^2 -\frac{j(j+1)}{r^2} \Big\}[r(R_{20}+R_{2\tilde{0}})]=0,
\end{equation}
\begin{equation}\label{eq4}
\Big\{\frac{d^2}{dr^2}+u^2 -\frac{j(j+1)}{r^2} \Big\}(R_{20}-R_{2\tilde{0}})=0,
\end{equation}
where $u^2=\big( \frac{E}{\hbar  c n}\big)^2-M^2$. Consequently, the solutions of Eqs. (\ref{eq1}, \ref{eq2}, \ref{eq3}, \ref{eq4}) are obtained as follows:
\begin{equation}\label{r2+}
R_{2+}+R_{2-}=\frac{N_{+}}{\sqrt{j(j+1)}}\frac{d(r\textnormal{j}_j)}{dr},
\end{equation}
\begin{equation}\label{r2-}
R_{2+}-R_{2-}=N_{0}(r\textnormal{j}_j),
\end{equation}
\begin{equation}\label{r20}
R_{20}+R_{2\tilde{0}}=N_{+} \textnormal{j}_j,
\end{equation}
\begin{equation}\label{r200}
R_{20}-R_{2\tilde{0}}=N_{-}r\textnormal{j}_j,
\end{equation}
where $N_0$ and $N_{\pm}$ are the integration constants, and $\textnormal{j}_j(ur)$ is the spherical Bessel function for integer $j$ which is the same solution of the radial part of the Helmholtz equation in spherical coordinates. We eliminate the other independent solution via the spherical Neumann function, because it diverges for our model when the argument of this solution $ur \ll 1$. 

The other helicities ($R_{1\pm}$, $R_{4\pm}$, $R_{10}$, $R_{40}$) corresponding to the classical polarization can be found by using the solutions in Eqs. (\ref{r2+}-\ref{r200}). However, we focus only on obtaining the resonance energies. Therefore, applying Eq. (\ref{boundary}) to Eqs. (\ref{eq2}, \ref{eq3}, \ref{eq4}), the approximate SR energies are calculated as follows:
\begin{equation}\label{massivereso}
E_{l,j} \approx \frac{hc}{2\pi b}\sqrt{j(j+1)+M^2b^2}\Big(1-\frac{1}{Q}\Big) , 
\end{equation}
by the approximation of $\frac{b-a}{a} \ll 1$ just as in Sect. \ref{sec:2}. To find an upper limit for the photon mass, Eq. (\ref{massivereso}) can be written as follows:
\begin{equation}\label{mass}
m_0 <  \frac{h \nu_j}{c^2}\sqrt{\Big( 1-\frac{1}{Q} \Big)^{-2} - \Big( \frac{f_j}{\nu_j} \Big)^{2}}.
\end{equation}
where $\nu$ is the modal frequency. Here, we see that the upper limit of the rest mass of the photon $m_0$ not only depends on $h \nu /c^2$ but also depends on the quality factor $Q$ and the natural Schumann resonances $f_j$ of the cavity, since we solve the massive spin-1 particle equation for the photon in the annular cavity.

\section{Concluding remarks}
\label{sec:4}
\hspace*{0.5cm} In this study, we solved the massless and the massive spin-1 particle equations in a spherical cavity under the condition of poorly conducting walls which means frequency-dependent quality factor. The resonance energy expression of the massless spin-1 particle in Eq. (\ref{masslessreso}) agrees well with the classical SR frequencies as we present in Table 1; but in this case, \(j \) is the quantum number of total angular momentum. Also, the quantum number of orbital angular momentum, \(l \), reads from \(l=0 \) in contrast to the classical mode number \citep{jackson}. In this sense, we summarize our results in Table 1 which are deduced from a statement that different $l$ values may occupy the same $j$ states; apparently the first fundamental SR may exist with the $j=1$ state produced from $l=0,1,2$ values. Furthermore, we may predict the SR by using the quality factor of the cavity in terms of the regularization parameter, $\epsilon_j$. Because the experimental results showing that the $Q$ factor is not an exact constant and changes slightly depending on the resonance  frequencies \citep{galejs2, Mushtak2002, Boldi2018}.

On the other hand, we show that the resonance energy expression of the massive spin-1 particle in Eq. (\ref{massivereso}) is equivalent to the massless case in Eq. (\ref{masslessreso}) in the \( m_{0}^2\rightarrow 0 \) limit. Therefore, an upper limit of the mass of the photon can be calculated from Eq. (\ref{mass}) as \(m_0 < 1.3 \times 10^{-50} \textnormal{ kg}$ by using the experimental first SR mode and the first mode averaged quality factor in Table 1 calculated from Eq. (\ref{quality}). The results in the historical and the recent studies try to set an upper limit on the photon rest mass are $m_0=\hbar/\lambda_0 c<4.2\times 10^{-49} \textnormal{ kg} $ where $\lambda_o > 8.3 \times 10^8$cm \citep{Kroll1971}, $m_0 < 1.6 \times 10^{-50} \textnormal{ kg}$ \citep{Williams1971}, $m_0 < 4.2 \times 10^{-50} \textnormal{ kg}$ \citep{Shao2017} and $m_0 < 3.9 \times 10^{-50} \textnormal{ kg}$ \citep{Bonetti2017}. Therefore, our limit on the photon rest mass reasonable is comparison with the given results.

\acknowledgments
We would like to thank the reviewers for important suggestions and Prof. Dr. Nuri Unal for the useful comments and discussions. This work was supported by the scientific research projects units of Akdeniz University.

\end{document}